\documentclass[twocolumn,10pt]{asme2ej}

\usepackage{epsfig} 
\usepackage{amsmath,amssymb}
\usepackage{graphicx}
\usepackage{subfig}
\usepackage{color}
\usepackage{multirow}
\usepackage{gensymb}
\usepackage{bm}
\graphicspath{{./figures/}}

\title{Computational Study of Pulmonary Flow Patterns 
    after Repair of Transposition of Great Arteries}

\author{Francesco Capuano
    \affiliation{
	Department of Industrial Engineering \\
    Universit\`a di Napoli Federico II \\
	Naples, Italy \\
    Email: francesco.capuano@unina.it
    }	
}

\author{Yue-Hin Loke
    \affiliation{
    Division of Cardiology \\
    Children's National Health System \\
	Washington DC, USA \\
	Email: yloke@childrensnational.org
    }
}

\author{Ileen Cronin
    \affiliation{
    Division of Cardiology \\
    Children's National Health System \\
	Washington DC, USA \\
	Email: icronin@childrensnational.org
    }
}

\author{Laura J. Olivieri
    \affiliation{
    Division of Cardiology \\
    The Sheikh Zayed Institute for \\
    Pediatric Surgical Innovation \\
    Children's National Health System \\  
	Washington DC, USA \\
	Email: lolivieri@childrensnational.org
    }
}

\author{Elias Balaras\thanks{Address all correspondence to this author.} 
    \affiliation{
    Department of Mechanical and Aerospace Engineering \\
    George Washington University \\
    Washington DC, USA \\
	Email: balaras@gwu.edu
    }
}

\begin{document}

\maketitle    

\begin{abstract}
{\it Patients that undergo the arterial switch operation (ASO) to repair transposition of great arteries (TGA) can develop abnormal pulmonary trunk morphology with significant long-term complications. In this study, cardiovascular magnetic resonance was combined with computational fluid dynamics to investigate the impact of the post-operative layout on the pulmonary flow patterns. Three ASO patients were analyzed and compared to a normal control. Results showed the presence of anomalous shear layer instabilities, vortical and helical structures, and turbulent-like states in all patients, particularly as a consequence of the unnatural curvature of the pulmonary bifurcation. Streamlined, mostly laminar flow was instead found in the healthy subject. These findings shed light on the correlation between the post-ASO anatomy and the presence of altered flow features, and may be useful to improve surgical planning as well as the long-term care of TGA patients.}
\end{abstract}

\begin{nomenclature}
\entry{AAo}{Ascending Aorta}
\entry{ASO}{Arterial Switch Operation}
\entry{CHD}{Congenital Heart Disease}
\entry{LPA}{Left Pulmonary Artery}
\entry{LV}{Left Ventricle}
\entry{MPA}{Main Pulmonary Artery}
\entry{PA}{Pulmonary Artery}
\entry{PT}{Pulmonary Trunk}
\entry{RPA}{Right Pulmonary Artery}
\entry{RV}{Right Ventricle}
\entry{TGA}{Transposition of Great Arteries}
\entry{WSS}{Wall Shear Stress}
\end{nomenclature}	

\section{Introduction} \label{sec:introduction}

\begin{figure*}
  \centering
  \includegraphics[width=.8\textwidth]{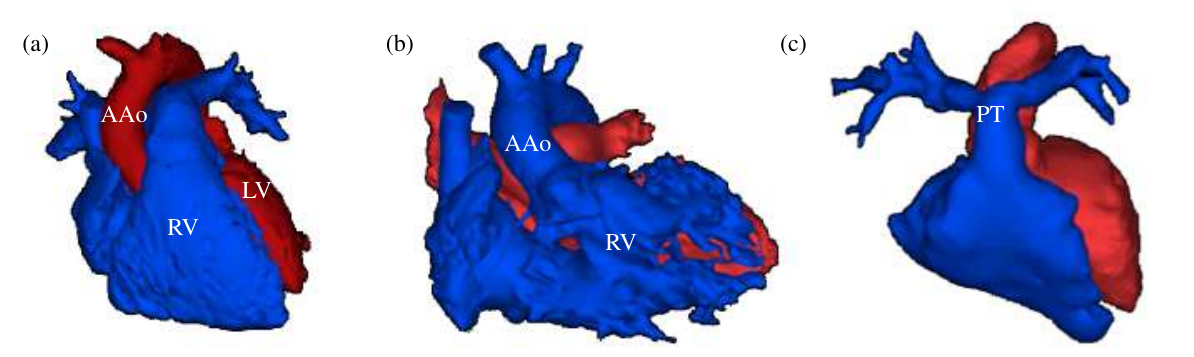}
  \caption{Comparison of healthy (a), pre-ASO (b) and post-ASO (c) arrangement of the great arteries shown by reconstructed patient-specific models. In the pre-ASO case (b), two separate circulatory systems are created: the right ventricle (RV) receives deoxygenated blood (blue) and pumps it back to the body via the ascending aorta (AAo), while the left ventricle (LV) continuously exchanges oxygen-rich blood (red) with the pulmonary circulation. After ASO (c), the coupling is restored although the pulmonary trunk (PT) wraps around the ascending aorta, leading to significant changes in morphology as compared to the physiological spiral anatomy (a). 
  \label{fig:comparison_intro}}
\end{figure*}

Dextro-transposition of great arteries (d-TGA) is a cyanotic defect accounting for 5\% to 7\% of all congenital heart disease (CHD) \cite{villafane2014d}. In d-TGA, the aorta and the pulmonary artery are connected to the right and left ventricle respectively, leading to a potentially fatal decoupling of systemic and pulmonary circulations, particularly in case of intact atrial septum \cite{shaher1964complete}. A corrective surgery is typically performed within 3-5 days from birth, and consists of the arterial switch operation (ASO) developed by Jatene \emph{et al.}\cite{jatene1976anatomic} in the mid-70s, which has now replaced the atrial switch procedure pioneered by Senning\cite{senning1959surgical} and Mustard\cite{mustard1964successful}. The ASO involves translocating the aorta and the pulmonary artery and re-implanting the coronary arteries into the neo-aorta \cite{fraser2017neonatal}. As a consequence of the so-called Lecompte maneuver, the pulmonary trunk is moved anterior to the aorta, with the pulmonary branches embracing the ascending aorta (AAo), in contrast with the physiological \textit{spiral} arrangement \cite{lecompte1981anatomic}. The pre- and post-ASO layouts of the great arteries, as compared to a healthy case, are shown in Fig.~\ref{fig:comparison_intro} by reconstructed three-dimensional models (see also Supplemental Video 1). The ASO with Lecompte maneuver can be considered a successful technique with a current rate of survival of 97\% at 25 years \cite{khairy2013cardiovascular}. However, several mid- and long-term sequelae have been reported (e.g., aortic regurgitation, pulmonary stenosis, limited aerobic capacity on exercise), requiring lifelong care of TGA patients and a relatively high rate of reintervention \cite{raju2013reoperation}.

\begin{figure*}
  \centering
  \includegraphics[width=.98\textwidth,trim={0 0.5cm 0 0.5cm},clip]{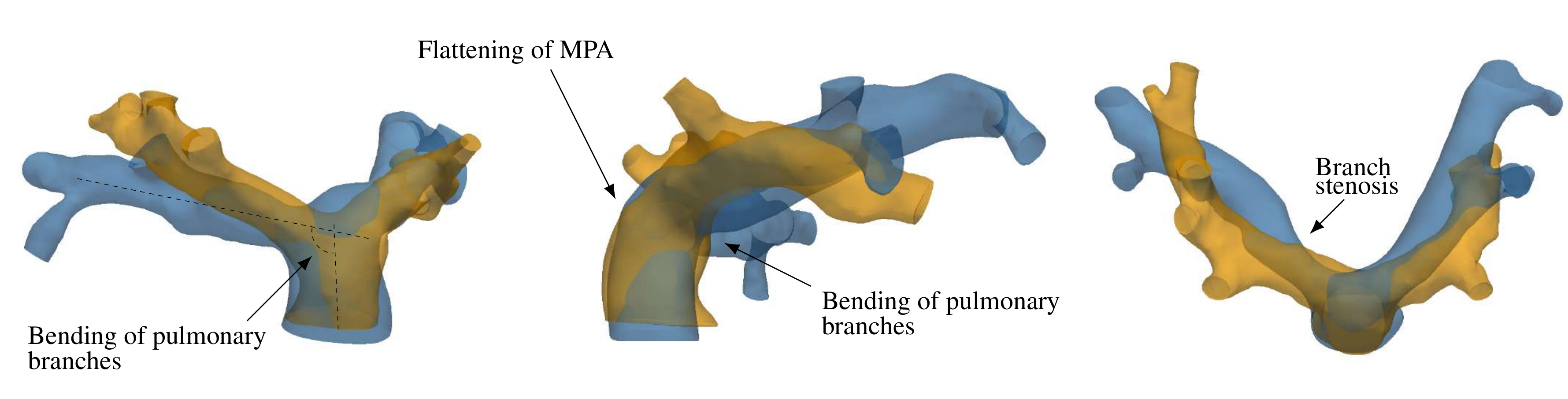}
  \caption{Main geometric differences between healthy (orange) and post-ASO (blue) pulmonary arteries. \label{fig:geometry}}
\end{figure*}

Some of the post-ASO complications are intimately related to the non-physiologic position of the great vessels after surgery. Being a low-pressure, highly-distensible system, the pulmonary arteries are particularly affected by this exceptional biomechanical environment, and their post-operative morphology can be profoundly different from the native anatomy. Pulmonary stenosis of various type and degree has been documented to appear in up to $60\%$ of the cases \cite{khairy2013cardiovascular}, supposedly due to the stretched course of the pulmonary branches around the AAo \cite{gutberlet2000arterial,morgan2016understanding}. The pulmonary trunk (PT) has been reported to be oval-shaped and flattened as a consequnce of being compressed between the AAo and the sternum \cite{ntsinjana20143d,geiger2014postoperative,massin1998growth}. An abnormal bending of the pulmonary arteries has been observed and described as the ``narrowing of the pulmonary arch window'' \cite{chiu2016concept}. These geometrical differences are summarized in Fig.~\ref{fig:geometry} by direct comparison between post-ASO and healthy three-dimensional models.

An important consequence of the abnormal pulmonary morphology is the distortion of blood flow patterns. Previous studies have shown a strong correlation between the post-ASO anatomy and the presence of altered flow features \cite{riesenkampff2014flow,markl2011comprehensive,sievers20164d}. In a pioneering computational fluid dynamics (CFD) analysis, Tang et al. \cite{tang2001comparison} demonstrated that the hemodynamic efficiency of the pulmonary arteries after ASO is generally reduced with respect to the normal anatomy. More recently, Geiger \emph{et al.}\cite{geiger2014postoperative} used 4D flow magnetic resonance imaging (MRI) to analyze post-operative hemodynamics in pediatric patients, as compared with healthy controls. They found increased vortical and helical patterns in the pulmonary arteries of TGA patients, and increased systolic flow velocities due to reduced vascular lumina. Abnormal vortex formation was also found by Rickers \emph{et al.}\cite{rickers2016lecompte} in a 4D flow investigation comparing the Lecompte maneuver to the spiral anamastosis, an alternative surgical procedure aiming to restore the spiral relationship of the great vessels. Despite the growing interest of the scientific community, there is still lack of knowledge regarding the onset mechanism and unsteady dynamics of the mentioned flow patterns, and their association to specific shape features of the post-ASO morphology is yet to be fully understood. Computational modeling is very appealing in this regard, as it can provide a wealth of information currently not obtainable with any other method \cite{morris2016computational}.

In this work, a MRI-driven, time- and space-resolved CFD analysis was conducted on three post-ASO pulmonary geometries and on one representative healthy control, with the aim of obtaining detailed information about the impact of the altered morphology on three-dimensional flow patterns. The results displayed complex flow features in all patients, in contrast with the streamlined velocity field found in the control case. A close inspection of the flowfields allowed to identify important geometric characteristics that are primarily responsible for the observed changes in flow physics. A thorough characterization of blood flow dynamics is supposed to have important clinical significance in terms of vascular remodeling, onset of atherosclerosis and right ventricle afterload \cite{morris2016computational}. With the oldest post-ASO patients being only in young adulthood, long-term consequences of the repair are not yet fully appreciated; these findings could thus be of help to inform lifelong care of such patients and suggest alternative surgical procedures.

\section{Methodologies} \label{sec:methods}

\subsection{Subjects dataset}

Cardiovascular magnetic resonance (CMR) imaging datasets of three patients with TGA status post ASO with branch pulmonary artery stenosis referred for cardiac catheterization were identified for this study. Catheterization and CMR data were obtained from the same sedation.  The three subjects were a 23 month old with right pulmonary artery stenosis, a 7 year old with main pulmonary artery stenosis and a 12 year old with right pulmonary artery stenosis. For normal controls, volunteer CMR datasets were identified. In the volunteer studies, non-contrast enhanced, respiratory-gated sequences were used to obtain 3D datasets of the pulmonary arteries. One normal control subject, a 5 year old female with no significant past medical history or peripheral branch pulmonary stenosis was eventually identified and selected. 

Institutional Review Board (IRB) approval was obtained for this study; all retrospective data was anonymized and a waiver of consent was obtained. Retrospectively-gated, through plane phase-encoded velocity mapping was performed using standard sequences, reconstructed to 30 phases per cardiac cycle with velocity encoding threshold set between 200 cm/s to 350 cm/s, depending on the peak velocity of accelerated blood flow as a result of branch stenosis.  Time-averaged flow rates in the MPA, LPA and RPA were derived from the phase velocity data and the pulmonary flow split was determined.  Review of phase velocity mapping demonstrated consistency in conservation of flow rates (the differences between RPA + LPA flow rate deviated from MPA flow rate by less than 10\%). We should note there were slightly abnormal flow split differences between the LPA and RPA for the ASO patients --- 56:44 in patient 1, 58:42 in patient 2 and 62:38 in patient 3. Normal controls are expected to be around 45:55 \cite{henk1998pulmonary}. The subjects' demographic information and main geometric and hemodynamic parameters are summarized in Table~\ref{tab:patients}.

\begin{table}
    \centering
    \begin{tabular}{ccccccc}
    Subject & Age & $A_\text{MPA}^{\text{avg}}$ & $Q_\text{MPA}^{\text{avg}}$ & $Q_\text{MPA}^{\text{peak}}$ & $Q_\text{LPA}^{\text{avg}}$ & $Q_\text{RPA}^{\text{avg}}$ \\[0.05cm]
    \hline
    Patient 1 & 2 & 1.96 & 37.2 & 113.3 & 20.0 & 15.5 \\
    Patient 2 & 7  & 2.14 & 64.5 & 196.9 & 35.1 & 24.8 \\
    Patient 3 & 12 & 2.80 & 51.1 & 205.7 & 40.7 & 25.6 \\
	Control & 5 & 2.73 & 74.4 & 257.1 & 32.8 & 39.9 \\    
    \hline
    \end{tabular}
    
    \caption{Demographic and hemodynamic information of the subjects. $A$ is the vessel area in cm$^2$, $Q$ is the flow rate in mL/s, the subscripts indicate the vessel, the superscript \textit{peak} refers to peak systole, \textit{avg} to the average over the cardiac cycle. The age is in years. \label{tab:patients}}
\end{table}

\subsection{Image Processing and 3D Model Construction}

\begin{figure*}
  \centering
  \includegraphics[width=.98\textwidth]{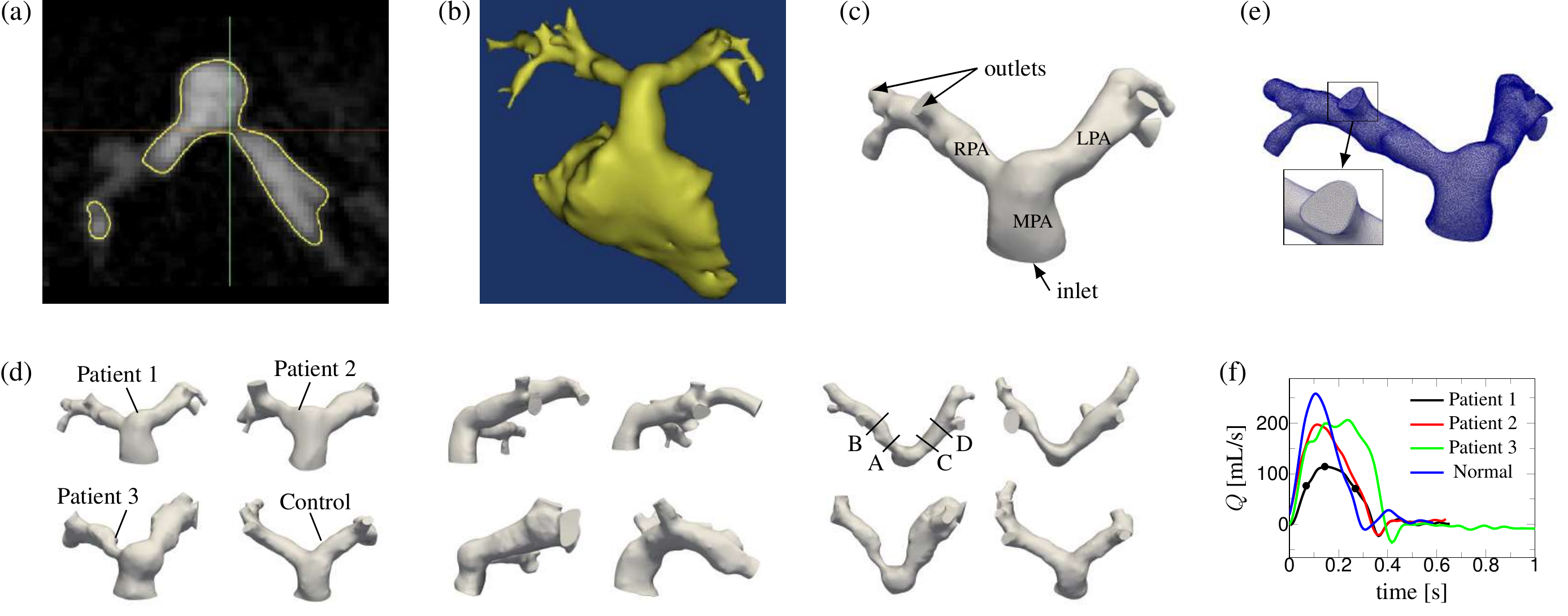}
  \caption{Segmentation, 3D modeling, meshing process and boundary conditions. (a) The blood pool representing the pulmonary arteries is identified and segmented by drawing a contour to create an overlying mask. (b) The segmented mask is then converted over to a 3D digital model in STL format. In this example, the right ventricle is also included as part of the initial model created. (c) 3D editing is performed to fine anatomic inlets and anatomic outlets. The outlets are cut as distally as allowed to create second-order branching. (d) From left to right: frontal, lateral and top view of the four models obtained via segmentation process. In the top view, the locations of the cut planes considered in the Results section are reported. (e) Exemplary mesh (Patient 1); the inset illustrates a detailed view of the surface mesh. (f) Inlet flow rates for the four models; shown are also characteristic points of the cardiac cycle considered in the Results section. \label{fig:segmentation}}  
\end{figure*}

For each dataset, the MR angiogram was used as a roadmap to build a 3D digital model of the pulmonary arteries via segmentation using image processing software (Mimics, Materialise, Belgium). Both automatic thresholding and manual methods were used to identify the blood pool of the pulmonary artery in each slice of the MR angiogram, allowing for creation of a 3D digital pulmonary artery model. The 3D model was then exported using the stereolithography (STL) file format.  This STL file was further edited and smoothed with 3D editing software (3-matic, Materialise, Belgium). Specific cut planes along the pulmonary architecture were made the defined anatomic inlets (main pulmonary artery) and anatomic outlets (right pulmonary artery and left pulmonary artery). To allow for flow curves to fully develop across the pulmonary architecture, the outlets were defined as distally as possible and thus cut across second-order branching. As a result, seven to eight outlets were defined for each model. An example of the overall segmentation and 3D modeling process is demonstrated in Fig.~\ref{fig:segmentation}(a-c), while Fig.~\ref{fig:segmentation}(d) reports front, lateral and top views of the four models analyzed in this study, clearly showing the morphological differences between the post-ASO geometries and the healthy case.

\subsection{Computational setup}

The subject-specific models were meshed using the simVascular open-source software \cite{updegrove2017simvascular}. All computations were started on a relatively coarse isotropic grid, which was successively refined until grid convergence of the time-averaged velocity field in selected points of the domain was achieved within $2\%$. The converged grid had approximately $35$ points along the MPA diameter and a maximum element size of about $450$ $\mu$m. The grid was further refined near the wall with a boundary layer mesh consisting of three elements normal to the outer surface, the smallest having a size of $200$ $\mu$m. On average, the models had $3$ million tetrahedral elements. An exemplary mesh (Patient 1) is shown in Fig.~\ref{fig:segmentation}(e).

The Navier-Stokes equations for incompressible flow were then solved for each discretized model, with proper boundary conditions, by means of the svSolver open-source package, which has been widely validated for cardiovascular simulations \cite{kung2011vitro,kung2014vitro}. This flow solver employs a robust stabilized finite-element method\cite{whiting2001stabilized} with backflow stabilization\cite{moghadam2011comparison}, and has been already applied in computational studies involving pulmonary arteries as well as pulmonary stenosis \cite{schiavazzi2015hemodynamic,tang2011three}. Blood was modeled as a Newtonian fluid with a density of $1050$ kg/m$^3$ and a viscosity of 0.035 Poise. The latter value was selected based on patients' hematocrit and well-established correlations. All the relevant scales of motion were solved and thus no turbulence model was employed.

Boundary conditions were obtained from the \textit{in-vivo} phase-contrast datasets.  The volumetric flow data was mapped onto the inlet face of the MPA as a discrete function of space and time. The flow data were available for a total of 30 points per cardiac cycle; the time-function was then discretely approximated by 10 Fourier harmonics to ensure periodicity and smooth the data sample.  The resulting flow waveform was spatially distributed at each time step using a plug flow profile, in accordance with phase-contrast data and literature experimental results \cite{morgan1998normal}. The flow rate waveforms employed for the models are reported in Fig.~\ref{fig:segmentation}(f).  At the outlets, the pressure was imposed by means of a resistive boundary condition \cite{vignon2006outflow}, and the values of the outlet resistances were computed as follows: i) the total resistances of the entire left and right branches were estimated (or computed from cath data, if available), and distributed to the branch vessels as to be inversely proportional to their cross-sectional area; ii) a simulation was then run and the LPA/RPA flow split was checked against MRI results; iii) if necessary, the outlet resistances were adjusted and the process was iterated until the flow split was captured within a 5\% accuracy. The entire process of choosing the outlet resistances typically converged in 2-3 iterations.  The walls of the vessels were assumed to be rigid and a no-slip condition was applied.

The simulations were run for several cardiac cycles, using a constant time-step size equal to $1 \times 10^{-4}$ s and up to 5 nonlinear iterations per time step. The time-step size was estimated on the basis of a CFL-like condition, by computing the maximum expected velocity in the domain and the size of the smallest grid element.  Simulation of one cardiac cycle took on average 8 hours when run in parallel on 16 processors at a high-end workstation.
 
\section{Results} \label{sec:results}

\subsection{Comparisons to clinical data}

\begin{figure*}
  \centering
  \includegraphics[height=5.5cm]{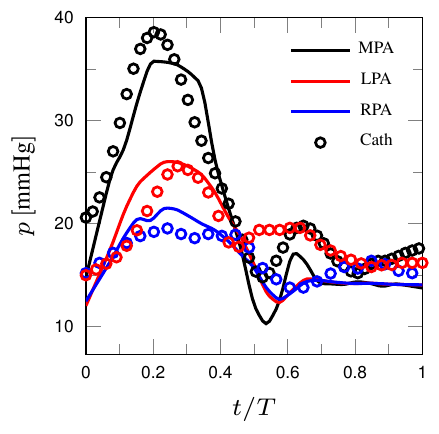}  
  \hspace{1cm}
  \includegraphics[height=5.5cm]{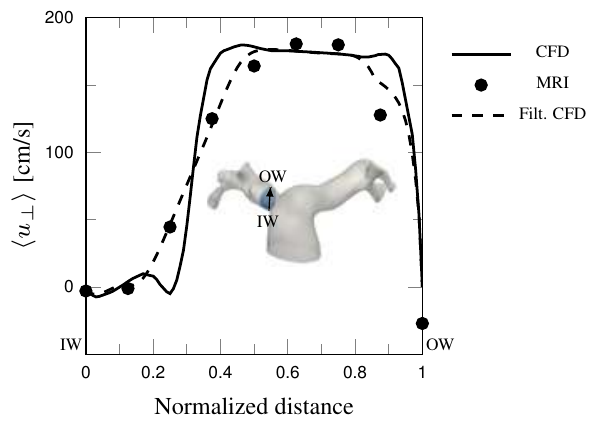}  
  \caption{Comparison of CFD results to clinical data. (left) Phase-averaged time signal of pressure in MPA, RPA and LPA; $T$ is the cardiac cycle period. (right) Phase-averaged streamwise velocity along a line moving from the inner wall (IW) to the outer wall (OW) of the RPA within the phase-contrast cut plane. \label{fig:CFD_clinical}}
\end{figure*}
Computational results were compared to available clinical data for validation purposes. The clinical dataset included both cardiac catheterization and phase-contrast MRI results, which were collected from the same sedation.  The latter enhances the confidence in the validation procedure as both velocity and pressure wave forms are available for analysis. A detailed comparison is presented for Patient 1. Figure~\ref{fig:CFD_clinical} (left) reports the time evolution of pressure at MPA, LPA and RPA during the cardiac cycle measured by the cardiac catheter and obtained by CFD.  In both cases, the pressure was phase-averaged for several cardiac cycles. The systolic phase and the mean pressure levels are satisfactorily captured by the computational model. The fact that the pressure levels found by CFD are in agreement with the cath data and compatible with the MRI-measured flow split ensures that the global hemodynamic balances are being correctly reproduced. On the other hand, higher-order features of the pressure signal could not be captured by CFD as a consequence of the modeling assumptions employed, namely i) the vessel compliance is not taken into account and therefore the CFD reports no phase lag between the MPA and LPA/RPA; ii) the discrepancy in pulse pressures is attributable to the purely resistive boundary conditions and iii) effects due to the valve closure are being only indirectly represented through the prescription of the flow rate at the MPA.  Further comments regarding the modeling limitations are reported in the Discussion section.

A comparison in terms of velocity profiles is shown in Fig.~\ref{fig:CFD_clinical} (right) at a representative location. The plot reports the phase-averaged streamwise velocity along a line lying on the RPA phase-contrast cut plane, moving from the inner wall to the outer wall of the proximal branch.  At a first glance the CFD results show significant differences when compared to the MRI data, which is however, due to the limited resolution of the latter.  When the CFD data are locally averaged to mimic the limited MRI resolution the agreement is very good.

The comparisons for all other cases was also satisfactory and a summary is given in Table~\ref{tab:verification}, in terms of systolic/diastolic pressures measured by cardiac catheterization and minimum/maximum velocity over the cut planes measured by phase-contrast MRI at peak systole (note that pressure data were not available for the healthy volunteer). The discrepancies between the pressures measured in the MPA by catheterization and the CFD results might be due to the simplified inlet condition, which is based on a flat velocity profile. This is particularly true for Patient 3, who presented with pulmonary valve stenosis and displayed a moderately distorted inlet flow shape.  For what concerns the velocity comparison, the peak velocities appear to be generally overpredicted by CFD. It has to be taken into account that the 3D models obtained via segmentation process are the result of a time average and are thus biased towards diastole. Therefore, on equal flow rate, the peak velocities obtained from the CFD results can be generally higher than those measured by MRI. For instance, the area change for the control case was rather high, $132\%$ for the LPA and $78\%$ for the RPA. It is also worth to mention that in most cases the peak velocity occurs in a very narrow region of the domain which most likely is not captured by the low-resolution phase-contrast data. It is however reassuring that the presence and magnitude of reverse flow was generally in good agreement between CFD and MRI.

\begin{table*}
    \centering
    \begin{tabular}{ccccccccccc}
    \multirow{4}{*}{Subject} & \multicolumn{6}{c}{Systolic/Diastolic Pressure [mmHg]} & \multicolumn{4}{c}{Min/Max Velocity [cm/s]} \\
    \cline{2-11}
              & \multicolumn{2}{c}{MPA} & \multicolumn{2}{c}{LPA} & \multicolumn{2}{c}{RPA} & \multicolumn{2}{c}{LPA} & \multicolumn{2}{c}{RPA} \\
              & Cath & CFD & Cath & CFD & Cath & CFD & MRI & CFD & MRI & CFD \\
    \hline
    Patient 1 & 35/12 & 35/12 & 25/14 & 24/12 & 17/11 & 18/12 & -19/189 & -54/189 & -45/180 & -43/198 \\
    Patient 2 & 23/9  & 31/9  & 23/9  & 25/9  & 16/9  & 16/9  & -26/160 & -38/198 & -14/193 & -45/220 \\
    Patient 3 & 38/12 & 43/12 & 33/13 & 33/12 & 24/12 & 23/12 & -97/166 & -42/166 & -29/157 & -49/236 \\
    Control   & n.a.  & 24/12 & n.a.  & 22/12 & n.a.  & 21/12 & -10/137 & -9/206  & -3/119  & 0/209 \\
    \hline
    \end{tabular}
    \caption{Comparison between phase-averaged CFD results and cath/MRI data. \label{tab:verification}}
\end{table*}

\subsection{Overview of Flow Features}

The flow features were qualitatively and quantitavely analyzed by means of several diagnostic parameters. Instantaneous visualizations of streamtraces were used to capture the overall behaviour of the flow.  The vortical and smaller flow structures were quantified via the Q-criterion \cite{huntQ}, which is the second invariant of the instantaneous velocity gradient tensor
\begin{equation}
Q = \dfrac{1}{2} \left(\| \boldsymbol{\Omega} \|^2 - \| \mathbf{S} \|^2  \right) \;,
\end{equation}
where $\mathbf{S}$ and $\boldsymbol{\Omega}$ are the symmetric and skew-symmetric decompositions of the velocity gradient respectively, and $\| \cdot \|$ is the Euclidean norm. Isosurfaces where $Q>0$ indicate regions where vorticity dominates over irrotational strain, and are thus indicative of coherent vortical structures. The vorticity itself, $\vec{\omega} = \nabla \times \vec{u}$, was visualized with the aim of detecting shear layer instabilities and areas of separated flow. Helical patterns, i.e., regions of the flow in which the velocity and the vorticity tend to be aligned, were identified by both streamtraces and by means of the normalized helicity density,
\begin{equation}
H = \dfrac{\vec{u} \cdot \vec{\omega}}{\| \vec{u} \| \| \vec{\omega} \| } \;,
\end{equation}
which describes the local value of the angle between the velocity and the vorticity vectors and varies between $-1$ and $1$ (left-handed and right-handed screw motion respectively).

\begin{figure*}
  \centering
  \includegraphics[width=.99\textwidth]{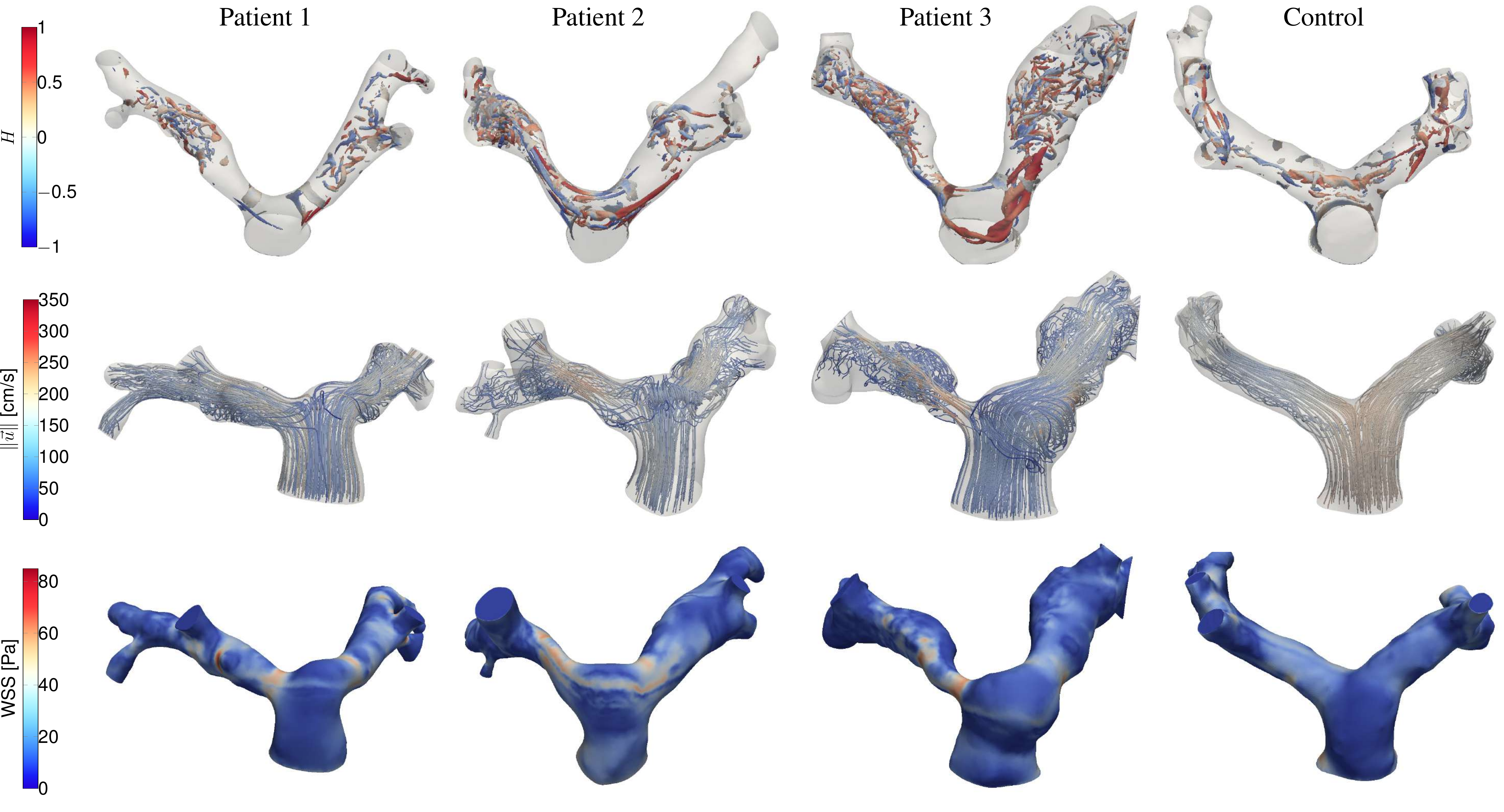}  
  \caption{Snapshot of the flow structure for the four subjects at peak systole. (top row) Vortical structures identified by isosurfaces of $Q=0.05Q_{\text{max}}$ coloured by normalized helicity. (middle row) streamlines coloured by velocity magnitude. (bottom row) wall shear stress. \label{fig:qcriterion}}
\end{figure*}

An instantaneous snapshot of the flow structure for all cases is reported in Fig.~\ref{fig:qcriterion}, which shows isosurfaces of $Q$ (coloured by normalized helicity), streamlines (coloured by velocity magnitude) and wall shear stress (WSS) at peak systole. In all patients, vortical structures are generated in the MPA, especially on its external wall, and are advected to the two branches, forming helical patterns due to the alignment between vorticity and velocity. A transition to a chaotic, turbulent-like state is observed in both RPA and LPA of all patients. The disturbance intensity is clearly higher as the stenosis severity increases (see the Q-criterion in the RPA from Patient 1 to Patient 3), although weakly disturbed flow is also seen in nominally non-stenotic vessels. Conversely, the flow in the healthy subject is mostly regular and streamlined throughout the entire cardiac cycle, except for one vortical motion forming on the internal side of the MPA and also producing a helix that protrudes predominantly in the RPA. High values of WSS are visible at the sites of the coherent vortical structures, as well as at the bifurcations. Local patches of high WSS are also present on the outer walls of stenotic vessels (especially in Patient 1 and Patient 3) as a result of the flow being redirected by the stenosis and impinging on the opposite wall. 

\begin{figure*}
  \centering
  \includegraphics[width=.95\textwidth]{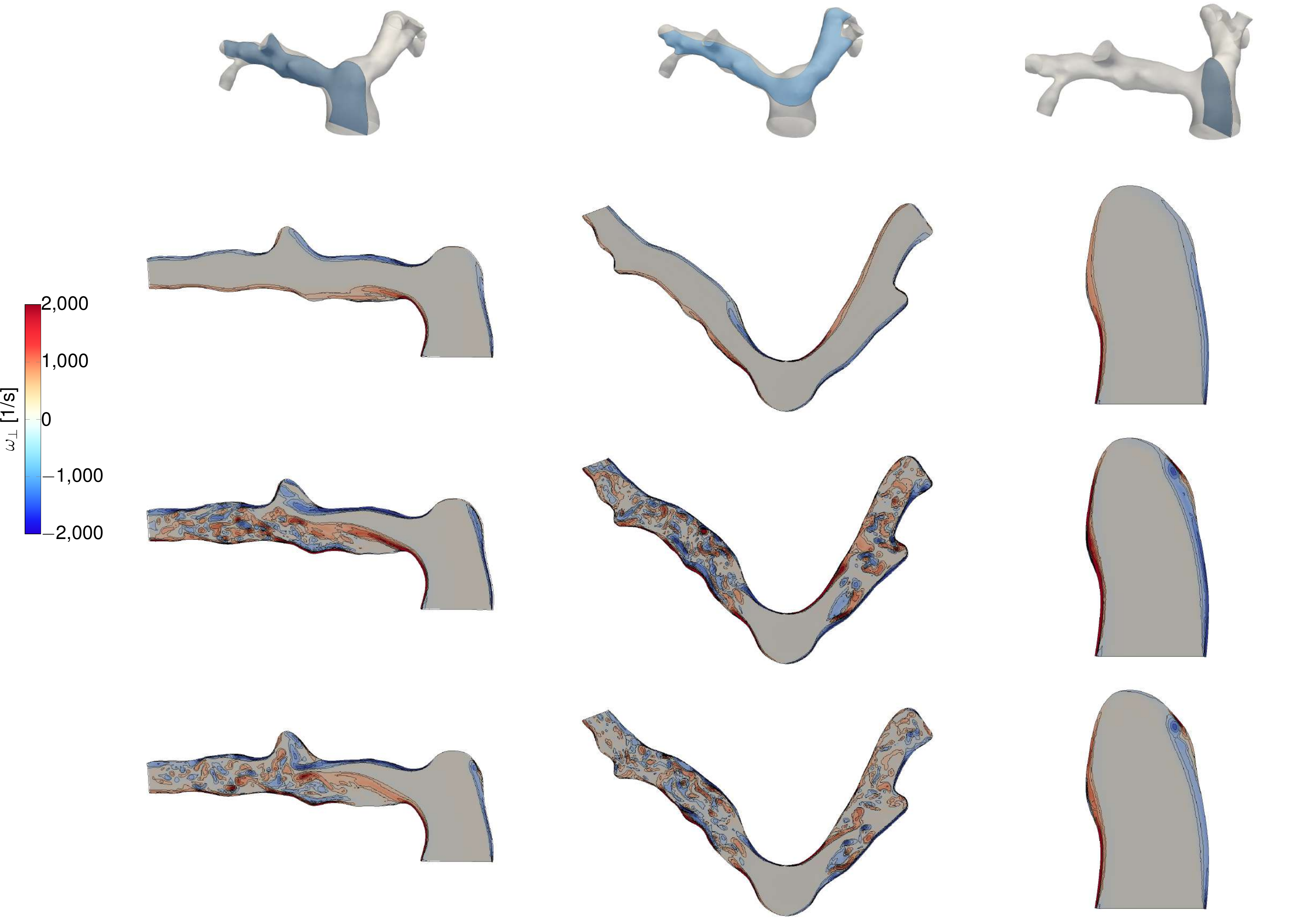}
  \caption{Normal-to-plane vorticity contours in the planes identified in the top row picture (Patient 1). From top to bottom: mid-acceleration, peak sysyole, and mid-deceleration. Contour lines are plotted for $\omega_\bot \in \left[ \pm 250,\: \pm 500,\: \pm 1000,\: \pm 1500,\: \pm 2000 \right]$. \label{fig:slices}}  
\end{figure*}

The transient mechanisms of vortex formation, flow separation and transition to a chaotic state in the post-ASO geometry are elucidated in Fig.~\ref{fig:slices} for one representative subject (Patient 1). Instantaneous contours of normal-to-plane vorticity are displayed in three slices and for three time instants: mid-acceleration, peak systole, and mid-deceleration (refer to Fig.~\ref{fig:segmentation}(d) for the position of these time instants along the MPA flow curve).  The first slice (left column) lies in a lateral plane and crosses the bend between MPA and RPA. The flow turns sharply within the bend and separates from the apex at the mid-acceleration; as a consequence, a long shear layer is formed which does not reattach to the inner wall. Near peak systole, the shear layer destabilizes and a turbulent-like flow is produced downstream, also persisting at mid-deceleration (see also Supplemental Video 2). A similar behavior takes place in the corresponding plane (not shown) crossing the LPA.
The mid column of Fig.~\ref{fig:slices} reports a slice in an axial view, crossing approximately the centerlines of the two branches. In the stenosed branch (right), a shear layer separation and destabilization mechanism occurs, similar to the one described for the MPA/RPA bend, while there is no significant evidence of flow instabilities occurring in this plane for the left branch. Therefore, the vortical structures seen in the LPA are the result of the apex-induced separation mechanism described above.
Finally, the right column shows a lateral slice crossing the MPA. It can be deduced that the vortical structure displayed in Fig.~\ref{fig:qcriterion} is the result of the wall shear layer detachment triggered by the MPA curvature, forming approximately during acceleration, and tending to weaken and disappear during the deceleration phase. However, while in the healthy case the flow detaches on the internal side of the MPA and reattaches shortly after, producing a well-defined recirculation area, in the post-ASO geometries the shear layer detaches predominantly from the external wall, destabilizes and rolls up during the systole. As the flow turns into the two branches, the velocity field advects and elongates the vortical structure, while aligning with vorticity. As a result, helical patterns are formed, as shown by the streamtraces in Fig.~\ref{fig:qcriterion}. The swirling flow is seen to travel within the proximal LPA and much more distally in the RPA of the healthy subject, whereas it quickly loses coherence in the proximal arteries of the patient, also due to the onset of a chaotic state. Additional information is provided by Supplemental Video 3 that shows the transient dynamics of these structures for Patient 2.

\begin{figure*}
  \centering
  \includegraphics[width=.9\textwidth]{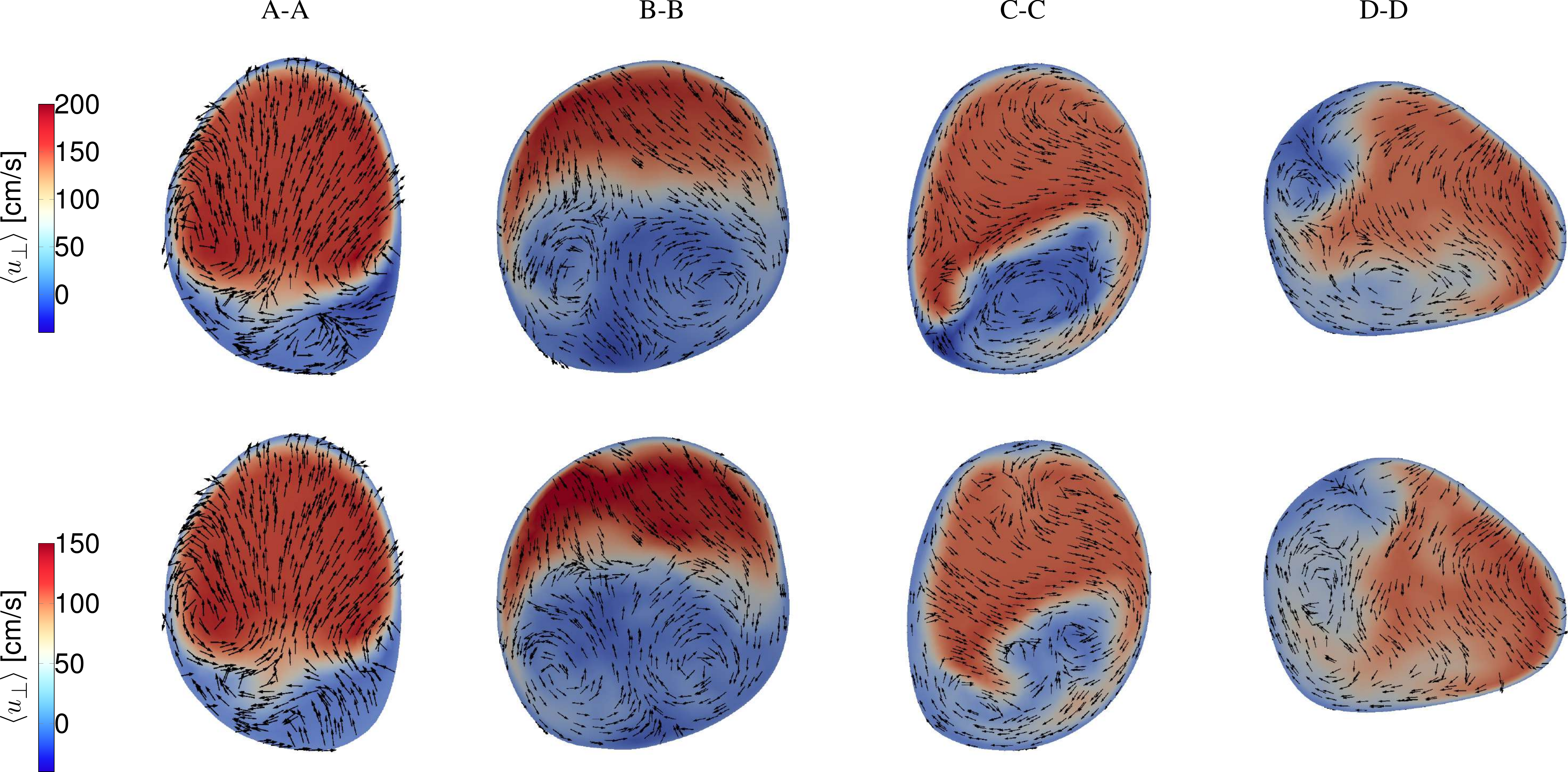}
  \caption{Phase-averaged velocity fields at peak systole (top row) and mid deceleration (bottom row) for Patient 1. Superimposed are in-plane phase-averaged velocity vectors. Cross sections are not to scale. \label{fig:phaseav_V}}  
\end{figure*}

\begin{figure*}
  \centering
  \includegraphics[width=.9\textwidth]{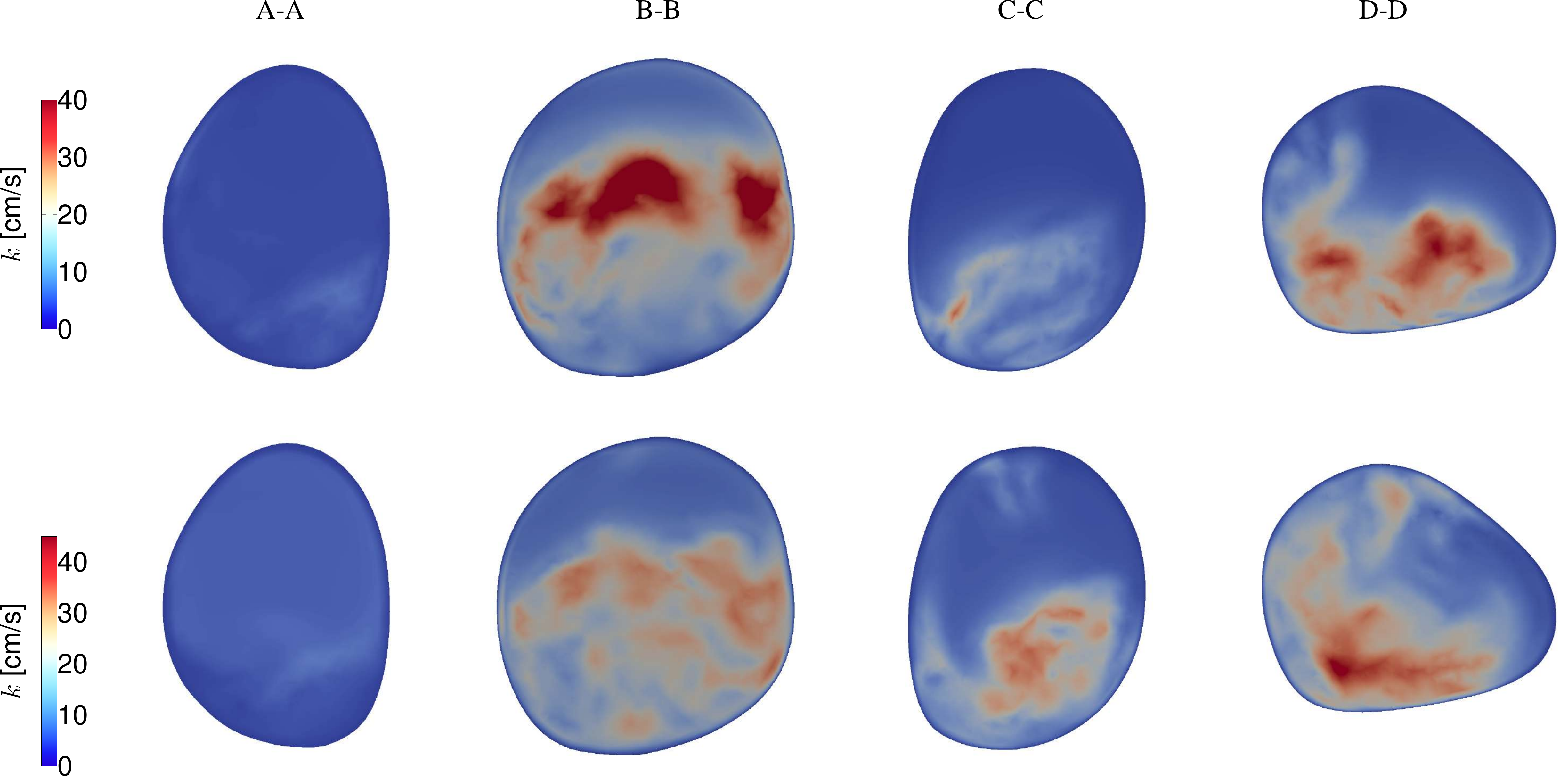}
  \caption{Phase-averaged square-root of turbulent kinetic energy at peak systole (top row) and mid deceleration (bottom row) for Patient 1. Cross sections are not to scale. \label{fig:phaseav_k}}  
\end{figure*}

Phase-averaged streamwise velocity fields are reported in Fig.~\ref{fig:phaseav_V} again for Patient 1 at slices perpendicular to the vessel centerlines and for two time instants: peak systole and mid-deceleration (all phase-averaged quantities, indicated as $\langle \; .\; \rangle$, were computed over six cardiac cycles). Instances of Dean-like flow patterns\cite{dean1928lxxii} are clearly visible from the superimposed in-plane velocity vectors. These secondary motions, in the form of counter-rotating vortices, generate within the branches as a result of the centripetal forces induced by the curvature \cite{berger1983flow}. However, the patterns are distorted due to the branch irregularities as well as to the curvature occurring in multiple planes. Phase averages of velocity fluctuations, $\vec{u}' = \vec{u} - \langle \vec{u} \rangle$, and the square root of the turbulent kinetic energy, as $k = \sqrt{\vec{u}' \cdot \vec{u}'/2}$, were also computed. Contours of $k$ are reported in Fig.~\ref{fig:phaseav_k} for the same cross sections. While the proximal LPA and RPA are almost free of disturbances, a remarkable amount of turbulent fluctuations (as high as $25 \%$ of the maximum streamwise velocity) is present at peak systole in the distal pulmonary arteries, being only slightly reduced at mid-deceleration. The turbulent stresses progress from the bottom wall towards the upper wall, according to the separation and shear-layer instability mechanisms described above. Note also that the apex-induced separation roughly divides a region of reverse flow from a high-speed jet. More distally in the two branches, the flow loses any coherent pattern and resembles a fully-developed turbulent flow.

\begin{figure*}
  \centering
  \includegraphics[width=.95\textwidth]{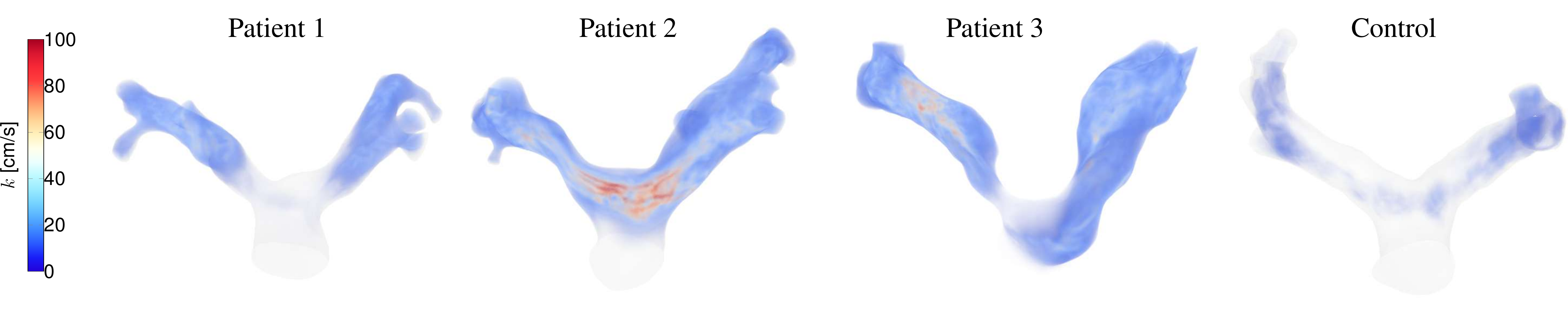}
  \caption{Phase-averaged square-root of turbulent kinetic energy at peak systole for all subjects. \label{fig:kpeak}}  
\end{figure*}

Volume-rendered plots of the phase-averaged square-root of turbulent kinetic energy at peak systole are shown for all subjects in Fig.~\ref{fig:kpeak}. In this representation, each point of the mesh is coloured according to the scalar intensity, and a certain level of translucency is added to allow for a fully three-dimensional visualization. The highest turbulent fluctuations occur in Patient 2, as a result of rather strong vortices detaching from the external wall of the MPA, due to the curvature-induced mechanism described above. In Patient 3, the generation of chaotic flow is predominantly caused by the stenosis, which is quite severe at the RPA: the wake of $k$ departing from the inner side of the stenotic region is clearly visible. Finally, it is worth to note that virtually no fluctuations (less than 2$\%$ of the streamwise velocity in the proximal arteries) occur in the healthy control.

\section{Discussion} \label{sec:discussion}

This study illustrated the detailed hemodynamics in the pulmonary arteries of TGA patients after ASO, as compared to a representative healthy subject. The three-dimensional CFD simulations reported significant topological changes in the flowfield of the repaired arteries, mainly as the result of the peculiar morphology induced by surgery. 

\begin{figure}
  \centering
  \includegraphics[width=.35\textwidth]{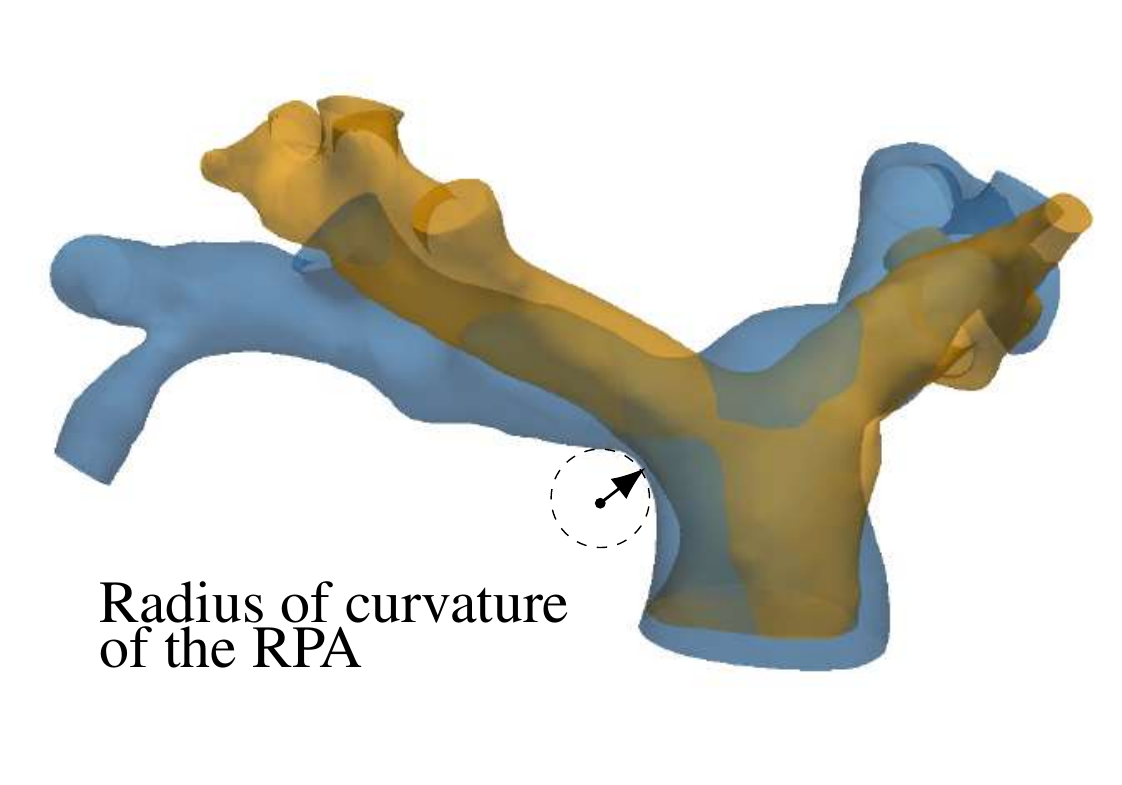}
  \caption{Illustrative definition of the radius of curvature of the bend between MPA and RPA. Blue: post-ASO geometry; orange: healthy model. \label{fig:curvature}}  
\end{figure}

The characteristics of the pulmonary bifurcation were determinant shape parameters. With reference to Fig.~\ref{fig:geometry}, the LPA/RPA angulation is defined as the angle between the MPA axis and the LPA/RPA axis in a lateral plane. In patients, the bending angle of the RPA was found to be significantly smaller compared to the control case: $103\degree$, $90\degree$ and $80\degree$ compared to $117\degree$. On the other hand, the LPA angles were similar: $114\degree$, $118\degree$ and $108\degree$ compared to $108\degree$ for the control case. In addition, a significant decrease in the radius of curvature of the bends was observed for patients in comparison to the healthy case, for both LPA and RPA. The curvature of each branch is defined in a lateral plane as the radius of the arc that approximates the inner wall curve, as this turns from the MPA to the two arterial branches, see Fig.~\ref{fig:curvature}. For the LPA, this radius was measured to be $2.7$ mm, $5.7$ mm and $11$ mm in patients and $19$ mm for the control; in the RPA, $4.6$ mm, $3.8$ mm and $5$ mm against $19$ mm of the healthy case. The decreased radius of curvature leads to an increase of the local Dean number, which is generally associated with complex flow features and transition to turbulence \cite{kuhnen2015subcritical,vester2016turbulent}. In the present cases, the combination of the above mentioned geometrical features promoted separation of the flow at the branches entrance in patients, recirculation areas, and subsequent flow instability and transition occurring distally in the LPA/RPA. The chaotic flow was also prompted by the relatively high Reynolds number at peak systole, which in the MPA was as high as $4000$, based on local flow velocity and vessel diameter.

Arterial bends and bifurcations are known to have a significant impact on hemodynamics \cite{ku1997blood}, although much of the previous work has focused mostly on the carotid bifurcation, due to its strong implications on atherosclerosis formation. Apparently, the critical role of the pulmonary bifurcation shape on flow patterns has been only partially recognized for post-operative CHD geometries. The influence of the pulmonary bifurcation morphology on hemodynamics was computationally explored by Chern and coworkers\cite{chern2012numerical,chern2008numerical} for patients with repaired tetralogy of Fallot (TOF). Similar to the results presented in this work, they found separated flow in the proximal pulmonary branches as a consequence of narrow bifurcation angles. More recently, Zhang \emph{et al.}\cite{zhang2016computational} analyzed the impact of the LPA angulation in TOF geometries again using CFD, and similarly concluded that acute angles (less than $100\degree$) are associated with adverse hemodynamic performance. 
While the curvature and the Dean number have been previously used by Hanna~\cite{hanna2005blood} to describe the pulmonary artery bifurcation, this is, to the authors' knowledge, the first study that recognizes the hemodynamic importance of the branch curvature for post-operative CHD geometries. Future work might explore the use of curvature measurements as potential markers of clinical relevance. Recently, curvature-related shape parameters of the aortic arch have been found to correlate well with indices of left ventricular function \cite{bruse2017successful}.

The presence of branch stenosis was also very impactful on hemodynamics. In the patients considered in this study, the narrowing occurred predominantly in an axial (top) plane crossing the centerlines of the LPA and RPA (see Fig.~\ref{fig:geometry}). The position of the ascending aorta subsequent to the Lecompte maneuver causes the bifurcation to be U-shaped in an axial plane, in contrast to the Y-shaped bifurcation of the normal pulmonary trunk.  A careful examination of the models (cf. Fig.~\ref{fig:comparison_intro} and Fig.~\ref{fig:segmentation}) shows clearly that the stenosis is presumably due to the compression of the pulmonary branches from the ascending aorta; this is especially visible for Patient 3. Due to the LPA and RPA embracing the aorta, the internal walls can present with a change in curvature even without any significant decrease of the lumen diameter (see for instance the LPA of Patient 2 and 3).  Previous studies also conjectured that the onset of pulmonary stenosis is supposedly related to the stretched course of the pulmonary arteries in conjunction with the compression of the pulmonary branch due to the high-pressure aortic dilatation \cite{gutberlet2000arterial,morgan2016understanding}. Typical features of stenotic flows were identified: flow acceleration, shear layer detachment and instability, and high values of wall shear stress both within the stenosis as well as on the opposite wall due to flow impingement \cite{lee2008direct,varghese2007direct}. Most importantly, in the patients considered in this study the abnormal flow split between left and right lung was primarily determined by the presence of the stenosis. On the other hand, the simulations showed that the chaotic flow downstream of the stenosis was the combination of both the apex-induced and the stenosis-induced instabilities.  An important consequence of this finding is that catheter-based reinterventions aimed to alleviate the stenosis could lead to only partial hemodynamic improvement.

Vortical and helical coherent structures were found in patients as well as in the control case. The dynamics of vortices and/or helical flow in the pulmonary arteries (and in the arterial system in general) is attracting increasing attention in literature, for the role of these flow patterns in both healthy and diseased vasculature, and for their potential clinical applications \cite{liu2015physiological}. However, research in this field is hampered by the fact that these structures are often localized in space and time, and therefore their detection is challenging for the current capabilities of imaging techniques, which typically provide phase-averaged and spatially-filtered (blurred) flow fields \cite{ku2002vivo}. In this regard, CFD can provide unrivalled spatial and temporal resolution and is thus a unique tool for the study of these complex flow features.  In all the present results, vortical flow generated firstly in the MPA. For the healthy subject, the simulations reported the presence of a swirling structure forming on the internal wall of the MPA and being spirally advected in the RPA. This is in accordance with the results of B\"{a}chler \textit{et al.}\cite{bachler2013assessment}, who found right-handed helical flow in the RPA in 15 out of 18 healthy volunteers. In the TGA patients, the flow separated especially on the external wall of the MPA, and either reattached shortly after, forming a recirculation region (as in Patient 1) or underwent a shear-layer instability, with vortex roll-up and shedding (Patients 2 and 3). The different behavior can be attributed to the modified MPA curvature taking place in the post-operative geometry. Visual inspection of the patients' models, compared to the healthy control (see Fig.~\ref{fig:segmentation}(d) and Fig.~\ref{fig:geometry}) shows that the MPA in the post-ASO cases is flattened and stands in a more upright posture, losing its natural curvature. Specifically, the external wall of the patients' MPA presents with sudden convex-to-concave changes, that can trigger flow instabilities. The healthy MPA walls have instead a monotonic curvature and smoothly extend into the LPA, preventing the flow to separate. The change in morphology described above is supposedly a result of the Lecompte maneuver, which leads the MPA to be ``straightened up'' by the ascending aorta. The pulmonary trunk position (in relation to the ascending aorta) is also known to play a role on the flow patterns in post-Lecompte cases \cite{geiger2014postoperative}. All the patients analyzed in this work had an anterior (central) position of the pulmonary trunk, a layout that was found to enhance vortical and helical flow \cite{geiger2014postoperative}, particularly when compared to the spiral arrangement \cite{rickers2016lecompte}. Based on similar fluid mechanics considerations, an alternative procedure aiming to restore the spiral relationship of the great vessels has been advocated by some authors \cite{tang2001comparison,chiu2010restoring}, although the Lecompte maneuver remains the standard clinical practice \cite{villafane2014d}. The correlation proposed in \cite{geiger2014postoperative} between the PT position and the vortical flow may be linked to the curvature of the pulmonary branches mentioned earlier in this section. Alleviating the RPA curvature to reduce flow separation and vortices might be an interesting input to design improved surgical procedures.

The modeling limitations of this work were primarily related to boundary conditions. 
The pulmonary valve was not included but rather a plug velocity profile based on the MRI-measured flow rate was prescribed at the MPA of each model.  This simplification is often made in numerical modeling of the pulmonary circulation\cite{kheyfets2013considerations}, and is justified by the results obtained by phase-contrast data, both in past literature work as well as in the present study, that indicate mostly flat velocity profiles at the MPA. Additional simulations (not reported here) using fully developed Poiseuille or Womersley solutions showed that the fundamental fluid dynamics features reported in the paper were insensitive to the inlet flow shape. Purely resistive outlet boundary conditions were employed. Although more sophisticated models (e.g., the Windkessel model) are capable to take into account the compliance of the pulmonary vasculature and to capture pulse pressures, this study was mainly concerned with the topology of the flow field, in terms of vorticity dynamics and transition to a chaotic state. As such the use of resistive outflow conditions is likely to have minimal effect on the flow features observed in this study.

Finally, a rigid wall assumption was used in this work. Presently, the inclusion of deformable wall models for the pulmonary circulation is very challenging, primarily due to the difficulties in obtaining reliable data of the vessel wall properties.
Although it is not trivial to assess the impact of the wall compliance on the results, prior studies \cite{perktold1995computer}, indicate that it primarily affects wall-related quantities (e.g., an over-estimation of the wall shear stress), and has a lesser impact on the overall flow patterns. Therefore, it is reasonable to expect that the results reported in the present study, especially with regards to the comparison between diseased and control models are representative.

\subsection{Concluding Remarks and Future Work}

This study reported for the first time detailed CFD results of the flowfields within the pulmonary arteries after correction of TGA with ASO and Lecompte maneuver. The onset mechanisms of abnormal flow patterns were elucidated and linked to certain shape features of the post-operative morphology. In particular, the narrow branch angulation and small radius of curvature of the LPA/RPA bends were found to contribute to flow separation and shear layer instability; the MPA wall curvature to the onset of helical patterns, and the degree of stenosis to flow acceleration and changes in wall shear stress. Overall, the flow characteristics found in the post-ASO geometries contribute to decrease the fluid mechanics efficiency and increase the local shear stresses, and thus may correlate with adverse post-operative progress and long-term complications, which are currently unknown due to the young age of post-ASO patients. A thorough knowledge of these phenomena is thus of great interest for surgical design and planning, and for the long-term care of TGA patients. Future morphometric studies are warranted to analyze the recurrency of the analyzed shape features in a larger TGA/ASO population, to identify the selective contribution of the geometrical parameters to power losses, as well as to take into consideration post-operative layouts occurring in different congenital heart diseases.

\section*{Conflict of Interest} 
The authors declare no conflict of interest.

\begin{acknowledgment}
The authors kindly acknowledge Russell R. Cross, MD.  F.C. was supported by Universit\`a degli Studi di Napoli Federico II and Compagnia di San Paolo under the STAR Program 2016.
\end{acknowledgment}

%
% ...

\bibliographystyle{asmems4}

\bibliography{bibliography}



\end{document}